\documentclass[sigconf,nonacm]{acmart}
\usepackage{subcaption}
\usepackage{hyperref}
\usepackage{multirow}
\usepackage{amsmath}
\usepackage{xcolor}
\usepackage{colortbl}
\usepackage{adjustbox}
\usepackage{fancyhdr}
\usepackage{setspace}
\usepackage{array,multirow}
\usepackage{makecell, tabularx}
\usepackage{color,soul}
\usepackage[most]{tcolorbox}
\usepackage{rotating, graphicx}
\usepackage{algpseudocode}
\usepackage{titlesec}
\usepackage{enumitem}

\setlist[itemize]{noitemsep, topsep=0pt}
\setlist[enumerate]{noitemsep, topsep=0pt}

\tcbset{
    left=8pt,
    right=8pt,
    enhanced,
    halign=center,
    colback=gray!16,
    colframe=black,
    boxrule=0.8pt,
    width=\dimexpr\textwidth\relax,
}

\usepackage{color}
\definecolor{darkred}{rgb}{0.55, 0.0, 0.0}
\definecolor{codegray}{rgb}{0.5,0.5,0.5}
\definecolor{mygreen}{RGB}{68,85,37}

\AtBeginDocument{%
  \providecommand\BibTeX{{%
    \normalfont B\kern-0.5em{\scshape i\kern-0.25em b}\kern-0.8em\TeX}}}

\copyrightyear{}
\acmYear{}
\setcopyright{acmcopyright}
\acmConference[] {}
\acmBooktitle{}
\acmPrice{}
\acmISBN{}
\acmDOI{}
\newcommand{\SparseP}{\emph{SparseP}}
\newcommand{\spmv}{SpMV}

\newcommand\fix[1]{\noindent{\color{black}{#1}}} 
\newcommand{\jgl}[1]{\textcolor{teal}{}}
\newcommand{\christina}[1]{\textcolor{blue}{}}

\begin{document}
\fancyhead{}
%\title{\SparseP: Towards Efficient Sparse Matrix Vector Multiplication on Real Processing-In-Memory Architectures}
\title{Towards Efficient Sparse Matrix Vector Multiplication \\ on Real Processing-In-Memory Systems}
\newcommand{\tsc}[1]{\textsuperscript{#1}} 
\newcommand{\affilETH}{\tsc{1}}
\newcommand{\affilNTUA}{\tsc{2}}
\newcommand{\affilUMA}{\tsc{3}}

\settopmatter{authorsperrow=1} 
\author{
{
Christina Giannoula\affilETH$^,$\affilNTUA\quad
Ivan Fernandez\affilETH$^,$\affilUMA\quad 
Juan Gómez-Luna\affilETH
}
}

\author{
{
Nectarios Koziris\affilNTUA\quad
Georgios Goumas\affilNTUA\quad
Onur Mutlu\affilETH 
}
}

\affiliation{
\institution{
      %\vspace{5pt}
      \affilETH ETH Z{\"u}rich \quad
      \affilNTUA National Technical University of Athens \quad
      \affilUMA University of Malaga
  }
   \vspace{0pt}
}
\renewcommand{\authors}{Christina Giannoula, Ivan Fernandez, Juan Gómez-Luna, Nectarios Koziris, Georgios Goumas, and Onur Mutlu}
\renewcommand{\shortauthors}{Christina Giannoula, et al.}


 \vspace{10pt}
\begin{abstract}

Sparse Matrix Vector Multiplication (\spmv) has been characterized as one of the most thoroughly studied scientific computation kernels, because it is a fundamental linear algebra kernel for important applications from the scientific computing, machine learning, and graph analytics domains. \spmv{} performs indirect memory references as a result of storing the sparse matrix in a compressed format, and irregular memory accesses to the input vector due to the sparsity pattern of the input matrix~\cite{Kanellopoulos2019SMASH,FacebookGraph,Goumas2009Performance}. Therefore, in commodity processor-centric systems, \spmv{} is a primarily memory-bandwidth-bound kernel for the majority of real sparse matrices, and is bottlenecked by data movement between memory and processors~\cite{Gomez2021Benchmarking,Gomez2021Analysis,Goumas2009Performance}.

One promising way to alleviate the data movement bottleneck is the Processing-In-Memory (PIM) paradigm~\cite{Gomez2021Benchmarking,Gomez2021Analysis,Giannoula2021SynCron,mutlu2020modern,devaux2019,Ghose2019Workload,Mutlu2019Processing,fernandez2020natsa,ahn2015scalable}. PIM moves computation close to application data by equipping memory chips with processing capabilities~\cite{mutlu2020modern,devaux2019,Mutlu2019Processing}. To provide large aggregate memory bandwidth for the in-memory processors, several manufacturers have already started to commercialize \textit{near-bank} PIM designs~\cite{Lee2021HardwareAA,Gomez2021Analysis,Gomez2021Benchmarking,devaux2019,SKhynix2022}. \textit{Near-bank} PIM designs tightly couple a PIM core with each DRAM bank, exploiting bank-level parallelism to expose high on-chip memory bandwidth of standard DRAM to processors. \fix{Three} \textit{real} near-bank PIM architectures are Samsung's FIMDRAM~\cite{Lee2021HardwareAA}, SK hynix's GDDR6-AiM~\cite{SKhynix2022} and the UPMEM PIM system~\cite{Gomez2021Analysis,Gomez2021Benchmarking,devaux2019}.

Most real near-bank PIM architectures~\cite{Lee2021HardwareAA,devaux2019,Gomez2021Analysis,Gomez2021Benchmarking,SKhynix2022} support several PIM-enabled memory chips connected to a host CPU via memory channels. Each memory chip comprises multiple low-power PIM cores with relatively low computation capability~\cite{Gomez2021Benchmarking,Gomez2021Analysis}, and each of them is located close to a DRAM bank~\cite{Lee2021HardwareAA,Gomez2021Analysis,Gomez2021Benchmarking,SKhynix2022}. Each PIM core can access data located on \fix{its} local DRAM bank, and typically there is no direct communication channel among PIM cores. \fix{Overall, near-bank PIM systems provide high levels of parallelism and very large memory bandwidth. As such, they are a very promising computing platform to accelerate memory-bound kernels.  Recent works leverage near-bank PIM architectures to provide high performance and energy benefits on bioinformatics~\cite{lavenier2020Variant,Gomez2021Benchmarking,Gomez2021Analysis}, skyline
computation~\cite{Zois2018Massively}, compression~\cite{Nider2020Processing} and neural network~\cite{Lee2021HardwareAA,Gomez2021Benchmarking,Gomez2021Analysis,Gu2020iPIM} kernels. A recent study~\cite{Gomez2021Analysis,Gomez2021Benchmarking} provides PrIM benchmarks~\cite{PrIMLibrary}, which are a collection of 16 kernels for evaluating near-bank PIM architectures. However, there is \emph{no} prior work to thoroughly study the widely used, memory-bound \spmv{} kernel} on a real PIM system.

Our work is the first to efficiently map the \spmv{} kernel on near-bank PIM systems, and understand its performance implications on a real-world PIM system. We make two key contributions. First, we design efficient \spmv{} algorithms to accelerate \fix{the \spmv{}} kernel in current and future PIM systems, while covering a wide variety of sparse matrices with diverse sparsity patterns. Second, we provide the first comprehensive analysis of \spmv{} on a real PIM architecture.  Specifically, we conduct our rigorous experimental analysis of \spmv{} kernels in the UPMEM PIM system, the first publicly-available real-world PIM architecture.

We present the \fix{freely and openly available} \SparseP{} library~\cite{SparsePLibrary} that includes 25 \spmv{} kernels for real PIM systems. \SparseP{} supports (1) the most popular compressed matrix formats (i.e., CSR,  COO, BCSR, BCOO formats), (2) a wide range of data types (i.e., 8-bit integer, 16-bit integer, 32-bit integer, 64-bit integer, 32-bit float and 64-bit float data types), (3) two types of well-crafted data partitioning techniques of the sparse matrix to PIM-enabled memory, (4) various load balancing schemes across PIM cores, (5) various load balancing schemes across threads of a multithreaded PIM core, and (6) three synchronization approaches among threads within multithreaded PIM core.

We conduct an extensive characterization \fix{and} analysis of \SparseP{} kernels on the UPMEM PIM system~\cite{Gomez2021Analysis,Gomez2021Benchmarking}. We analyze the \spmv{} execution (1) \fix{using} one single multithreaded PIM core, (2) \fix{using} thousands of PIM cores, and (3) comparing its performance and energy consumption with that achieved on conventional processor-centric CPU and GPU systems. Our extensive evaluation provides programming recommendations for software designers, and suggestions and hints for hardware and system designers of future PIM systems.

We highlight our most significant recommendations for PIM software designers:

\begin{enumerate} [noitemsep, leftmargin=*, topsep=0pt]
    \item \textit{Design algorithms that provide high load balance across threads of a multithreaded PIM core in terms of computations, loop control iterations, synchronization points and memory accesses.} In \spmv{}, we find that when the parallelization scheme used causes \emph{high} disparity in the non-zero elements/blocks/rows processed across threads of a PIM core, or the number of lock acquisitions/lock releases/DRAM memory accesses performed across threads, %\spmv{} 
    performance severely degrades in low-area PIM cores with relatively low computation capabilities~\cite{Gomez2021Analysis,Gomez2021Benchmarking}.
    \item \textit{Design compressed data structures that can be effectively partitioned across DRAM banks, with the goal of providing high computation balance across PIM cores.} We observe that the compressed matrix format used to store the input matrix in \spmv{} determines the data partitioning across DRAM banks of PIM-enabled memory, thereby affecting the load balance across PIM cores with corresponding performance implications.
    \item \textit{Design \textit{adaptive} algorithms that trade off computation balance across PIM cores for lower data transfer costs to PIM-enabled memory, and adapt the \fix{software strategies} to the particular patterns of each input given, as well as the characteristics of the PIM hardware.} Our analysis demonstrates that the best-performing \spmv{} execution on the UPMEM PIM system is achieved using algorithms that (i) trade off computation for lower data transfer costs, and (ii) select the load balancing strategy and data partitioning policy based on the particular sparsity pattern of the input matrix and the characteristics of the underlying PIM hardware. %(e.g., the number of PIM cores, off-chip memory bus bandwidth, microarchitecture of the host CPU cores).
\end{enumerate}

We highlight our most significant suggestions for PIM hardware and system designers:

\begin{enumerate} [noitemsep, leftmargin=*, topsep=0pt]
    \item \textit{Provide low-cost synchronization support and hardware support to enable concurrent memory accesses by multiple threads to the local DRAM bank to increase parallelism in a multithreaded PIM core.} For instance, fine-grained locking approaches in \spmv{} to increase parallelism in critical sections do not improve performance over coarse-grained approaches. This is because concurrent DRAM accesses performed by multiple threads are serialized by the UPMEM PIM hardware. To improve parallelism, subarray level parallelism~\cite{Kim2012Case} or multiple DRAM banks per PIM core could be supported in the PIM hardware, along with lightweight synchronization schemes for PIM cores~\cite{Giannoula2021SynCron}.
    \item \textit{Optimize the broadcast collective operation in data transfers from main memory to PIM-enabled memory to minimize overheads of copying the input data into all DRAM banks in the PIM system.} When the sparse matrix is horizontally partitioned across PIM cores and the whole input vector is copied into the DRAM bank of \emph{each} PIM core, \spmv{} cannot scale up to a large number of PIM cores. This is because it is severely limited by data transfer costs to broadcast the input vector into each DRAM bank of PIM-enabled DIMMs. Such data transfers incur high overheads, because they take place via the narrow off-chip memory bus.
    \item \textit{Optimize the gather collective operation \textit{at DRAM bank granularity} for data transfers from PIM-enabled memory to the host CPU to minimize overheads of retrieving the output results.} When the sparse matrix is split in 2D tiles, each of them is assigned to each PIM core, \spmv{} is severely limited by data transfers to retrieve results for the output vector from DRAM banks of PIM-enabled memory. This is due to two reasons: (i) PIM cores create a large number of partial results that need to be gathered from PIM-enabled memory to the host CPU via the narrow memory bus in order to assemble the final output vector, and (ii) the \fix{current implementation of the} UPMEM PIM system has the limitation that the transfer sizes from/to all DRAM banks involved in the same parallel transfer need to be the same, and therefore a large amount of padding with empty bytes is performed in such \spmv{} kernels. 
    \item \textit{Design high-speed communication channels and optimized libraries for data transfers to/from thousands of DRAM banks of PIM-enabled memory.} We find that \spmv{} execution on the memory-centric UPMEM PIM system achieves a much higher fraction of the machine’s peak performance (on average 51.7\% for the 32-bit float data type), compared to that on processor-centric CPU and GPU systems. However, its end-to-end performance is still significantly limited by data transfer overheads on the narrow memory bus. Thus, the software stack of real PIM systems needs to be enhanced with fast data transfers to/from PIM-enabled memory modules, and/or the PIM hardware \fix{needs to be enhanced to} support efficient direct communication among PIM cores~\cite{Seshadri2013RowClone,Chang2016LISA,Wang2020Figaro,Rezaei2020NoM}.
\end{enumerate}

For more information about our thorough characterization on the \spmv{} PIM execution, results, insights and the \fix{open-source} \SparseP{} software package~\cite{SparsePLibrary}, we refer the reader to the full version of the paper~\cite{Giannoula2022SparsePPomacs,Giannoula2022SparseParXiv}. We hope that our work can provide valuable insights to programmers in the development of efficient sparse linear algebra kernels and other irregular kernels from different application domains tailored for real PIM systems, and enlighten architects and system designers in the development of future memory-centric computing systems. The \SparseP{} software package is publicly and freely available at \textcolor{blue}{\href{https://github.com/CMU-SAFARI/SparseP}{https://github.com/CMU-SAFARI/SparseP}}. %to enable further research on \spmv{} in current and future PIM systems.

\end{abstract}

%\keywords{high-performance computing, HPC, sparse matrix-vector multiplication, SpMV, SpMV library, multicore, processing-in-memory, near-data processing, memory systems, data movement bottleneck, DRAM, benchmarking, real-system characterization, workload characterization}

\maketitle

\vspace{-7pt}
\bibliographystyle{ACM-Reference-Format}
%\setstretch{0.97}
\bibliography{references}

\end{document}